\documentclass[envcountsame]{llncs}

\usepackage{amsmath}
\usepackage{amssymb}
\usepackage{graphicx}
\usepackage[dvips,ps,all]{xy}

\title{Extending the Overlap Graph for Gene Assembly in Ciliates}
\author{Robert Brijder \and Hendrik Jan Hoogeboom}

\institute{Leiden Institute of Advanced Computer Science, Universiteit Leiden,\\
Niels Bohrweg 1, 2333 CA Leiden, The Netherlands,\\
\email{rbrijder@liacs.nl}}

\def\blfootnote{\xdef\@thefnmark{}\@footnotetext}
\long\def\symbolfootnote[#1]#2{\begingroup%
\def\thefootnote{\fnsymbol{footnote}}\footnote[#1]{#2}\endgroup}

\newenvironment{Theorem}{\begin{theorem}}{\end{theorem}}

\newenvironment{Lemma}{\begin{lemma}}{\end{lemma}}
\newenvironment{Corollary}{\begin{corollary}}{\end{corollary}}
\newenvironment{Remark}{\begin{remark}\rm}{\qed\end{remark}}

\newenvironment{Example}{\begin{example}\rm}{\end{example}}
\newenvironment{Definition}{\begin{definition}\rm}{\end{definition}}
\newenvironment{Proof}{\begin{proof}}{\qed\end{proof}}

\begin{document}
\newcommand{\bridge}{\mathrm{bridge}}
\newcommand{\erase}{\mathrm{erase}}
\newcommand{\snrdom}{\mathrm{snrdom}}
\newcommand{\edom}{\mathrm{dom}}
\newcommand{\dom}{\mathrm{dom}}
\newcommand{\prem}{\mathrm{rem}}
\newcommand{\rf}{r \! f}
\newcommand{\pset}[1]{\|#1\|}
\newcommand{\merge}{\mathrm{merge}}
\newcommand{\RGVertL}[1]{I_{#1}}
\newcommand{\RGVertR}[1]{I'_{#1}}
\newcommand{\dedge}{\ar@{-}}
\newcommand{\redge}{\ar@2{-}}
\newcommand{\redgr}[1]{\mathcal{R}_{#1}}
\newcommand{\rgrem}[2]{\mathcal{R}_{\prem_{#2}(#1)}}
\newcommand{\pcgr}{\mathcal{PC}}
\newcommand{\setmultgr}{{\sf MGr}}
\newcommand{\settwoedgegr}{{\sf 2EGr}}
\newcommand{\overlap}{\mathcal{G}}
\newcommand{\eoverlap}{\mathcal{G}}
\newcommand{\snr}{{\bf snr}}
\newcommand{\sspr}{{\bf sspr}}
\newcommand{\ssdr}{{\bf ssdr}}
\newcommand{\gnr}{{\bf gnr}}
\newcommand{\sgpr}{{\bf sgpr}}
\newcommand{\gnrset}{\mathrm{Gnr}}
\newcommand{\sgprset}{\mathrm{sGpr}}
\newcommand{\rmpar}[1]{[#1]}
\newcommand{\rmundir}[1]{[[#1]]}
\newcommand{\remm}{\mathrm{rm}}
\newcommand{\range}{\mathrm{rng}}

\renewcommand{\emptyset}{\varnothing}
\bibliographystyle{plain}

\maketitle

\begin{abstract}
Gene assembly is an intricate biological process that has been
studied formally and modeled through string and graph rewriting
systems. Recently, a restriction of the general (intramolecular)
model, called simple gene assembly, has been introduced. This
restriction has subsequently been defined as a string rewriting
system. We show that by extending the notion of overlap graph it is
possible to define a graph rewriting system for two of the three
types of rules that make up simple gene assembly. It turns out that
this graph rewriting system is less involved than its corresponding
string rewriting system. Finally, we give characterizations of the
`power' of both types of graph rewriting rules. Because of the
equivalence of these string and graph rewriting systems, the given
characterizations can be carried over to the string rewriting
system.
\end{abstract}

\section{Introduction}
Gene assembly is a highly involved process occurring in one-cellular
organisms called ciliates. Ciliates have two both functionally and
physically different nuclei called the micronucleus and the
macronucleus. Gene assembly occurs during sexual reproduction of
ciliates, and transforms a micronucleus into a macronucleus. This
process is highly parallel and involves a lot of splicing and
recombination operations -- this is true for the stichotrichs group
of ciliates in particular. During gene assembly, each gene is
transformed from its micronuclear form to its macronuclear form.
\symbolfootnote[0]{This research was supported by the Netherlands
Organization for Scientific Research (NWO) project 635.100.006
`VIEWS'.}

Gene assembly has been extensively studied formally, see
\cite{GeneAssemblyBook}. The process has been modeled as either a
string or a graph rewriting system
\cite{Equiv_String_Graph_1,Equiv_String_Graph_2}. Both systems are
`almost equivalent', and we refer to these as the general model. In
\cite{ModelSimpleOps} a restriction of this general model has been
proposed. While this model is less powerful than the general model,
it is powerful enough to allow each known gene \cite{Ciliate_DB} in
its micronuclear form to be transformed into its macronuclear form.
Moreover this model is less involved and therefore called the simple
model. The simple model was first defined using signed permutations
\cite{ModelSimpleOps}, and later proved equivalent to a string
rewriting system \cite{DBLP:conf/fct/BrijderLP07}. The graph
rewriting system of the general model is based an overlap graphs.
This system is an abstraction from the string rewriting system in
the sense that certain local properties within the strings are lost
in the overlap graph. Therefore overlap graphs are not suited for
the simple gene assembly model. In this paper we show that by
naturally extending the notion of overlap graph we can partially
define simple gene assembly as a graph rewriting system. These
extended overlap graphs form an abstraction of the string model, and
is some way easier to deal with. This is illustrated by
characterizing the power of two of the three types of recombination
operations that make up simple gene assembly. While this
characterization is based on extended overlap graphs, due to its
equivalence, it can be carried over to the string rewriting system
for simple gene assembly.

\section{Background: Gene Assembly in Ciliates}
\begin{figure}
\begin{center}
\resizebox{\textwidth}{!}{\input{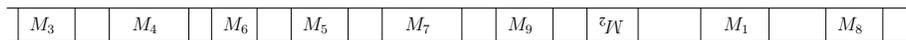}}
\caption{The structure of the MIC gene encoding for the actin
protein in sterkiella nova.}\label{intro_fig_mic_gene}
\end{center}
%\rotatebox{180}{$M_2$}
\end{figure}

In this section we very briefly describe the process of gene
assembly. For a detailed account of this process we refer to
\cite{GeneAssemblyBook}. Gene assembly occurs in a group of
one-cellular organisms called ciliates. A characterizing property of
ciliates is that they have two both functionally and physically
different nuclei called the micronucleus (MIC) and the macronucleus
(MAC). Each gene occurs both in the MIC and in the MAC, however they
occur in very different forms in the MIC and the MAC. The MIC form
of a gene consists of a number of DNA segments $M_1, \ldots,
M_{\kappa}$, called MDSs, which occur in some fixed permutation on a
chromosome. The MDSs are separated by non-coding DNA segments.
Moreover, each MDS can occur inverted, i.e. rotated 180 degrees. For
example, the gene in MIC form encoding for the actin protein in
ciliate sterkiella nova is given in Figure~\ref{intro_fig_mic_gene}
(see \cite{Biology_of_Ciliates,Ciliate_DB}). Notice that $M_2$
occurs inverted.

\begin{figure}
\begin{center}
\resizebox{\textwidth}{!}{\input{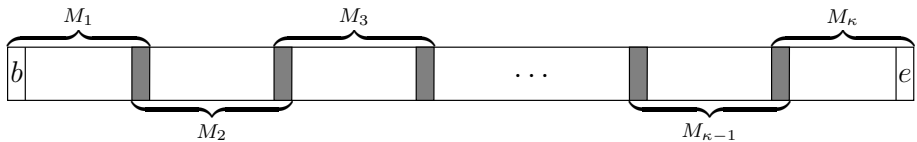}} \caption{The
structure of a MAC gene consisting of $\kappa$
MDSs.}\label{intro_fig_mac_gene}
\end{center}
\end{figure}

In the MAC form of the gene, the MDSs occur as a sequence $M_1,
\cdots, M_{\kappa}$ where each two consecutive MDSs overlap, see
Figure~\ref{intro_fig_mac_gene}. These shaded area's represent the
overlapping segments and are called \emph{pointers}. Moreover, there
are two sequences denoted by $b$ and $e$, which occur on $M_1$ and
$M_{\kappa}$ respectively, that indicate the beginning and ending of
the gene. The sequences $b$ and $e$ are called \emph{markers}. The
process of gene assembly transforms the MIC into the MAC, thereby
transforming each gene in the MIC form to the MAC form. Hence, for
each gene the MDSs are `sorted' and put in the right orientation
(i.e., they do not occur inverted). This links gene assembly to the
well-known theory of sorting by reversal \cite{Extended_paper}.

It is postulated that there are three types of recombination
operations that cut-and-paste the DNA to transform the gene from the
MIC form to the MAC form. These operations are defined on pointers,
so one can abstract from the notion of MDSs by simply considering
the MIC gene as a sequence of pointers and markers, see
Figure~\ref{fig_seq_pointers} corresponding to the gene in MIC form
of Figure~\ref{intro_fig_mic_gene}. The pointers are numbered
according to the MDS they represent: the pointer on the left (right,
resp.) of MDS $M_i$ is denoted by $i$ ($i+1$, resp.). Pointers or
markers that appear inverted are indicated by a bar: hence pointers
$2$ and $3$ corresponding to MDS $M_2$ appear inverted and are
therefore denoted by $\bar 2$ and $\bar 3$ respectively. In the
general model the markers are irrelevant, so in that case only the
sequence of pointers is used.

\begin{figure}
\resizebox{\textwidth}{!}{\input{Genome_with_pointers_large.pstex_t}}
\caption{Sequence of pointers and markers representing the gene in
MIC form.} \label{fig_seq_pointers}
\end{figure}

\section{Legal Strings with Markers}

For an arbitrary finite alphabet $A$, we let $\bar A = \{ \bar a
\mid a \in A \}$ with $A \cap \bar A = \emptyset$. We use the `bar
operator' to move from $A$ to $\bar A$ and back from $\bar A$ to
$A$. Hence, for $p \in A \cup \bar A$, $\bar {\bar {p}} = p$. For a
string $u = x_1 x_2 \cdots x_n$ with $x_i \in A$, the \emph{inverse}
of $u$ is the string $\bar u = \bar x_n \bar x_{n-1} \cdots \bar
x_1$. We denote the empty string by $\lambda$.

We fix $\kappa \geq 2$, and define the alphabet $\Delta =
\{2,3,\ldots,\kappa\}$ and the alphabet $\Pi = \Delta \cup \bar
\Delta$. The elements of $\Pi$ are called \emph{pointers}. For $p
\in \Pi$, we define $\pset{p}$ to be $p$ if $p \in \Delta$, and
$\bar{p}$ if $p \in \bar{\Delta}$, i.e., $\pset{p}$ is the
`unbarred' variant of $p$.
%The \emph{domain} of a string $u \in \Pi^*$ is $\dom(u) = \{
%\pset{p} \mid \mbox{$p$ occurs in $u$} \}$.
%
%We define $\edom(u) = \dom(u) \cup \{m\}$.
A \emph{legal string} is a string $u \in \Pi^*$ such that for each
$p \in \Pi$ that occurs in $u$, $u$ contains exactly two occurrences
from $\{p,\bar p\}$.

Let $M = \{b,e\}$ with $\Delta \cap \{b,e\} = \emptyset$. The
elements of $M$ are called \emph{markers}. We let $\Xi = \Delta \cup
\{b,e\}$, and let $\Psi = \Xi \cup \bar \Xi$. We define the morphism
$\remm:\Psi^* \rightarrow \Pi^*$ as follows: $\remm(a)=a$, for all
$a \in \Pi$, and $\remm(m) = \lambda$, for all $m \in M \cup
\bar{M}$. We say that a string $u \in \Psi^*$ is an \emph{extended
legal string} if $\remm(u)$ is a legal string and $u$ has one
occurrence from $\{b, \bar{b}\}$ and one occurrence from $\{e,
\bar{e}\}$. We fix $m \not\in \Psi$ and define for each $q \in M
\cup \bar{M}$, $\pset{q} = m$.

An extended legal string represents the sequence of pointers and
markers of a gene in MIC form. Hence, the extended legal string
corresponding to Figure~\ref{fig_seq_pointers} is $3 4 4 5 6 7 5 6 7
8 9 e \bar 3 \bar 2 b 2 8 9$. The legal string corresponding to this
figure is $3 4 4 5 6 7 5 6 7 8 9 \bar 3 \bar 2 2 8 9$ (without the
markers). Legal strings are considered in the general model since
markers are irrelevant there.

The \emph{domain} of a string $u \in \Psi^*$ is $\edom(u) = \{
\pset{p} \mid \mbox{$p$ occurs in $u$} \}$. Note that $m \in
\edom(v)$ for each extended legal string $v$. Let $q \in \edom(u)$
and let $q_1$ and $q_2$ be the two occurrences of $u$ with
$\pset{q_1} = \pset{q_2} = q$. Then $q$ is \emph{positive} in $u$ if
exactly one of $q_1$ and $q_2$ is in $\Xi$ (the other is therefore
in $\bar{\Xi}$). Otherwise, $q$ is \emph{negative} in $u$.

\begin{Example}
String $u = 24b4\bar e \bar 2$ is an extended legal string since
$\remm(u) = 244\bar2$ is a legal string. The domain of $u$ is
$\edom(u) = \{m,2,4\}$. Now, $m$ and $2$ are positive in $u$, and
$4$ is negative in $u$.
\end{Example}

Let $u = x_1 x_2 \cdots x_n$ be an (extended) legal string with $x_i
\in \Xi$ for $1 \leq i \leq n$, and let $p \in \edom(u)$. The
\emph{$p$-interval} of $u$ is the substring $x_i x_{i+1} \cdots x_j$
where $i$ and $j$ with $i < j$ are such that $\pset{x_i} =
\pset{x_j} = p$.

Next we consider graphs. A signed graph is a graph $G =
(V,E,\sigma)$, where $V$ is a finite set of \emph{vertices}, $E
\subseteq \{\{x,y\} \mid x,y \in V, x\not=y\}$ is a set of
(undirected) \emph{edges}, and $\sigma: V \rightarrow \{+,-\}$ is a
\emph{signing}, and for a vertex $v \in V$, $\sigma(v)$ is the
\emph{sign} of $v$. We say that $v$ is \emph{negative} in $G$ if
$\sigma(v) = -$, and $v$ is \emph{positive} in $G$ if $\sigma(v) =
+$. A signed directed graph is a graph $G = (V,E,\sigma)$, where the
set of edges are directed $E \subseteq V \times V$. For $e =
(v_1,v_2) \in E$, we call $v_1$ and $v_2$ \emph{endpoints} of $e$.
Also, $e$ is an edge \emph{from} $v_1$ \emph{to} $v_2$.

\section{Simple and General String Pointer Rules}

Gene Assembly has been modeled using three types of string rewriting
rules on legal strings. These types of rules correspond to the types
of recombination operations that perform gene assembly. We will
recall the string rewriting rules now -- together they form the
string pointer reduction system, see
\cite{Equiv_String_Graph_1,GeneAssemblyBook}. The string pointer
reduction system consists of three types of reduction rules
operating on legal strings. For all $p,q \in \Pi$ with $\pset{p}
\not = \pset{q}$:
\begin{itemize}
\item
the \emph{string negative rule} for $p$ is defined by
$\textbf{snr}_{p}(u_1 p p u_2) = u_1 u_2$,
\item
the \emph{string positive rule} for $p$ is defined by
$\textbf{spr}_{p}(u_1 p u_2 \bar p u_3) = u_1 \bar u_2 u_3$,
\item
the \emph{string double rule} for $p,q$ is defined by
$\textbf{sdr}_{p,q}(u_1 p u_2 q u_3 p u_4 q u_5) = $ \\ $u_1 u_4 u_3
u_2 u_5$,
\end{itemize}
where $u_1,u_2,\ldots,u_5$ are arbitrary (possibly empty) strings
over $\Pi$.

We now recall a restriction to the above defined model. The
motivation for this restricted model is that it is less involved but
still general enough to allow for the successful assembling of all
known experimental obtained micronuclear genes  \cite{Ciliate_DB}.
The restricted model, called \emph{simple} gene assembly, was
originally defined on signed permutations, see
\cite{ModelSimpleOps,DBLP:journals/fuin/LangilleP06}. The model can
also be defined as string rewriting rules (in an equivalent way) as
done for the general model above. This model is defined and proven
equivalent in \cite{DBLP:conf/fct/BrijderLP07}, and we recall it
here. It turns out that it is necessary to use extended legal
strings adding symbols $b$ and $e$ to legal strings.

The simple string pointer reduction system consists of three types
of reduction rules operating on \emph{extended} legal strings. For
all $p,q \in \Pi$ with $\pset{p} \not = \pset{q}$:
\begin{itemize}
\item
the \emph{string negative rule} for $p$ is defined by $\snr_{p}(u_1
p p u_2) = u_1 u_2$ as before,
\item
the \emph{simple string positive rule} for $p$ is defined by
$\sspr_{p}(u_1 p u_2 \bar p u_3) = u_1 \bar u_2 u_3$, where $|u_2| =
1$, and
\item
the \emph{simple string double rule} for $p,q$ is defined by
$\ssdr_{p,q}(u_1 p q u_2 p q u_3) = u_1 u_2 u_3$,
\end{itemize}
where $u_1$, $u_2$, and $u_3$ are arbitrary (possibly empty) strings
over $\Psi$. Note that the string negative rule is not changed, and
that the simple version of the string positive rule requires $|u_2|
= 1$, while the simple version of the string double rule requires
$u_2 = u_4 = \lambda$ (in the string double rule definition).

\begin{Example}
Let $u = 5 \bar 2 4 4 \bar 5 3 \bar 6 2 6 b 3 \bar e$ be an extended
legal string. Then within the simple string pointer reduction system
only $\snr_{4}$ and $\sspr_{\bar 6}$ are applicable to $u$. We have
$\sspr_{\bar 6}(u) = 5 \bar 2 4 4 \bar 5 3 \bar 2 b 3 \bar e$.
Within the string pointer reduction system also $\textbf{spr}_{5}$
and $\textbf{spr}_{\bar 2}$ are applicable to $u$. We will use $u$
(in addition to a extended legal string $v$, which is defined later)
as a running example.
\end{Example}

A composition $\varphi = \rho_n \ \cdots \ \rho_2 \ \rho_1$ of
string pointer rules $\rho_i$ is a \emph{reduction} of (extended)
legal string $u$, if $\varphi$ is applicable to (i.e., defined on)
$u$. A reduction $\varphi$ of legal string $u$ is \emph{successful}
if $\varphi(u) = \lambda$, and a reduction $\varphi$ of extended
legal string $u$ is \emph{successful} if $\varphi(u) \in \{be, eb,
\bar e \bar b, \bar b \bar e\}$. A successful reduction corresponds
to the transformation using recombination operations of a gene in
MIC form to MAC form. It turns out that not every extended legal
string has a successful reduction using only simple rules -- take
e.g. $2 3 4 \bar 2 \bar 3 \bar 4$.

\begin{Example}
In our running example, $\varphi = \sspr_{\bar 3} \ \sspr_{2} \
\sspr_{5} \ \snr_{4} \ \sspr_{\bar 6}$ is a successful reduction of
$u$, since $\varphi(u) = \bar b \bar e$. All rules in $\varphi$ are
simple.
\end{Example}

\section{Extended Overlap Graph}

The general string pointer reduction system has been made more
abstract by replacing legal strings by so-called overlap graphs, and
replacing string rewriting rules by graph rewriting rules. The
obtained model is called the graph pointer reduction system.
Unfortunately, this model is not fully equivalent to the string
pointer reduction system since the string negative rule is not
faithfully simulated. Also, overlap graphs are not suited for a
graph model for simple gene assembly. We propose an extension to
overlap graphs that allows one to faithfully model the string
negative rule and the simple string positive rule using graphs and
graph rewriting rules. First we recall the definition of overlap
graph.

\begin{Definition}
The \emph{overlap graph} for (extended) legal string $u$ is the
signed graph $(V,E,\sigma)$, where $V = \dom(u)$ and for all $p,q
\in \dom(u)$, $\{p,q\} \in E$ iff $q \in \dom(p')$ and $p \in
\dom(q')$ where $p'$ ($q'$, resp.) is the $p$-interval
($q$-interval) of $u$. Finally, for $p \in \edom(u)$, $\sigma(p) =
+$ iff $p$ is positive in $u$.
\end{Definition}

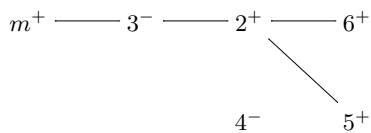
\begin{figure}
$$
\xymatrix{
m^+ \ar@{-}[r] & 3^- \ar@{-}[r] & 2^+ \ar@{-}[r] \ar@{-}[dr] & 6^+ \\
& & 4^- & 5^+
}
$$
\caption{The overlap graph of $u$ from Example~\ref{ex1_overlapgr}.}
\label{fig1_overlapgr}
\end{figure}

\begin{Example} \label{ex1_overlapgr}
Consider again extended legal string $u = 5 \bar 2 4 4 \bar 5 3 \bar
6 2 6 b 3 \bar e$. Then the overlap graph $\eoverlap_u$ of $u$ is
given in Figure~\ref{fig1_overlapgr}.
\end{Example}

We say that $p,q \in \dom(u)$ \emph{overlap} if there is an edge
between $p$ and $q$ in the overlap graph of $u$. We now define the
extended overlap graph.

\begin{Definition}
The \emph{extended overlap graph} for (extended) legal string $u$ is
the signed directed graph $(V,E,\sigma)$, denoted by $\eoverlap_u$,
where $V = \edom(u)$ and for all $p,q \in \edom(u)$, there is an
edge $(q,p)$ iff $q$ or $\bar q$ occurs in the $p$-interval of $u$.
Finally, for $p \in \edom(u)$, $\sigma(p) = +$ iff $p$ is positive
in $u$.
\end{Definition}

%\begin{figure}
%$$
%\xymatrix{
%%
%m^+ \ar@/^0.2pc/[r] & 3^- \ar@/^0.2pc/[r] \ar@/^0.2pc/[l] & 2^+ \ar@/^0.2pc/[r] \ar@/^0.2pc/[dr] \ar@/^0.2pc/[l] & 6^+ \ar@/^0.2pc/[l] \\
%%
%& & 4^- \ar@/^0.2pc/[r] \ar@/_0.2pc/[r] \ar@/_0.2pc/[u]
%\ar@/^0.2pc/[u] & 5^+ \ar@/^0.2pc/[ul]
%%
%}
%$$
%\caption{The extended overlap graph of $u$ from
%Example~\ref{ex1_extended_overlapgr}.}
%\label{fig1_extended_overlapgr}
%\end{figure}

Notice first that between any two (different) vertices $p$ and $q$
we can have the following possibilities:
\begin{enumerate}
\item There is no edge between them. This corresponds to $u =
u_1 p u_2 p u_3 q u_4 q u_5$ or $u = u_1 q u_2 q u_3 p u_4 p u_5$
for some (possibly empty) strings $u_1, \ldots, u_5$ and possibly
inversions of the occurrences of $p$ and $q$ in $u$.
\\
\item
There are exactly two edges between them, which are in opposite
direction. This corresponds to the case where $p$ and $q$ overlap in
$u$.
%,like in the specification of the string double rule.
%
\\
\item There is exactly one edge between them. If there is an edge
from $p$ to $q$, then this corresponds to the case where $u = u_1 q
u_2 p u_3 p u_4 q u_5$ for some (possibly empty) strings $u_1,
\ldots, u_5$ and possibly inversions of the occurrences of $p$ and
$q$ in $u$.
\end{enumerate}

As usual, we represent two directed edges in opposite direction
(corresponding to case number two above) by one undirected edge. In
the remaining we will use this notation and consider the extended
overlap graph as having two sets of edges: undirected edges and
directed edges. In general, we will call graphs with a special
vertex $m$ and having both undirected edges and directed edges
$\emph{simple marked graphs}$.
%, where each two edges do not have exactly the same endpoints,

\begin{figure}
$$
\xymatrix{
m^+ \ar@{-}[r] & 3^- \ar@{-}[r] & 2^+ \ar@{-}[r] \ar@{-}[dr] & 6^+ \\
& & 4^- \ar[r] \ar[u] & 5^+
}
$$
\caption{The extended overlap graph of $u$ from
Example~\ref{ex1_extended_overlapgr}.}
\label{fig2_extended_overlapgr}
\end{figure}
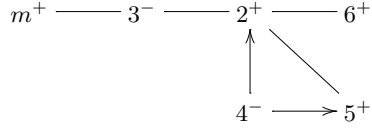

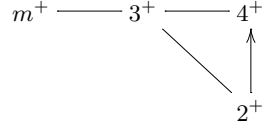
\begin{figure}
$$
\xymatrix{
m^+ \ar@{-}[r] & 3^+ \ar@{-}[r] \ar@{-}[dr] & 4^+ \\
& & 2^+ \ar[u]
}
$$
\caption{The extended overlap graph of $v$ from
Example~\ref{ex1_extended_overlapgr}.}
\label{fig3_extended_overlapgr}
\end{figure}

\begin{Example} \label{ex1_extended_overlapgr}
Consider again extended legal string $u = 5 \bar 2 4 4 \bar 5 3 \bar
6 2 6 b 3 \bar e$. Then the extended overlap graph $\eoverlap_u$ of
$u$ is given in Figure~\ref{fig2_extended_overlapgr}. Also, the
extended overlap graph of $v = \bar 4 2 3 \bar 2 4 \bar e \bar 3 b$
is given in Figure~\ref{fig3_extended_overlapgr}.
\end{Example}

The undirected graph obtained by removing the directed edges is
denoted by $\rmpar{\eoverlap_u}$. This is the `classical' overlap
graph of $u$, cf. Figures~\ref{fig1_overlapgr} and
Figure~\ref{fig2_extended_overlapgr}. On the other hand, the
directed graph obtained by removing the undirected edges is denoted
by $\rmundir{\eoverlap_u}$. This graph represents the proper nesting
of the $p$-intervals in the legal string.

\section{Simple Graph Rules}
We will now define two types of rules for simple marked graphs
$\gamma$. Each of these rules transform simple marked graph of a
certain form into another simple marked graph. We will subsequently
show that in case $\gamma$ is the extended overlap of a legal
strings, then these rules faithfully simulate the effect of the
$\snr$ and $\sspr$ rules on the underlying legal string.

\begin{Definition} \label{def_gnr_sgpr}
Let $\gamma$ be a simple marked graph. Let $p$ be any vertex of
$\gamma$ not equal to $m$.
\begin{itemize}
\item
The \emph{graph negative rule} for $p$, denoted by $\gnr_p$, is
applicable to $\gamma$ if $p$ is negative, there is no undirected
edge $e$ with $p$ as an endpoint, and there is no directed edge from
a vertex \emph{to} $p$ in $\gamma$. The result is the simple marked
graph $\gnr_p(\gamma)$ obtained from $\gamma$ by removing vertex $p$
and removing all edges connected to $p$. The set of all graph
negative rules is denoted by $\gnrset$.
\\
\item
The \emph{simple graph positive rule} for $p$, denoted by $\sgpr_p$,
is applicable if $p$ is positive, there is exactly one undirected
edge $e$ with $p$ as an endpoint, and there is no directed edge from
a vertex \emph{to} $p$ in $\gamma$. The result is the simple marked
graph ${\bf sgpr}_p(\gamma)$ obtained from $\gamma$ by removing
vertex $p$, removing all edges connected to $p$, and flipping the
sign of the other vertex $q$ of $e$ (i.e. changing the sign of $q$
to $+$ if it is $-$ and to $-$ if it is $+$). The set of all simple
graph positive rules is denoted by $\sgprset$.
\end{itemize}
These rules are called simple graph pointer rules.
%Note: the applicability is easier using the formal definition of extended overlap graph
\end{Definition}

\begin{Remark}
The $\sgpr$ rule is much simpler than the ${\bf gpr}$ for
`classical' overlap graphs. One does not need to compute the `local
complement' of the set of adjacent vertices. Obviously, this is
because the simple rule allows only a single pointer in the
$p$-interval.
\end{Remark}

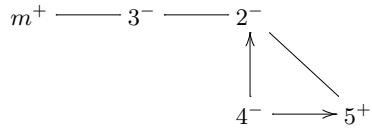
\begin{figure}
$$
\xymatrix{
m^+ \ar@{-}[r] & 3^- \ar@{-}[r] & 2^- \ar@{-}[dr] \\
& & 4^- \ar[r] \ar[u] & 5^+
}
$$
\caption{The simple marked graph $\gnr_4(\eoverlap_u)$.}
\label{fig_gnr_extended_overlapgr}
\end{figure}

\begin{Example} \label{ex_apply_graph_rules}
Rules $\gnr_4$ and $\sgpr_6$ are the only applicable rules on the
simple marked graph $\gamma = \eoverlap_u$ of
Figure~\ref{fig2_extended_overlapgr}. Simple marked graph
$\sgpr_6(\gamma)$ is depicted in
Figure~\ref{fig_gnr_extended_overlapgr}.
\end{Example}

Similar as for strings, a composition $\varphi = \rho_n \ \cdots \
\rho_2 \ \rho_1$ of graph pointer rules $\rho_i$ is a
\emph{reduction} of simple marked graph $\gamma$, if $\varphi$ is
applicable to (i.e., defined on) $\gamma$. A reduction $\varphi$ of
$\gamma$ is \emph{successful} if $\varphi(\gamma)$ is the graph
having only vertex $m$ where $m$ is negative. For $S \subseteq
\{\gnrset,\sgprset\}$, we say that $\gamma$ is \emph{successful in
$S$} if there is a successful reduction of $\gamma$ using only graph
pointer rules from $S$.

\begin{Example}
In our running example, $\varphi = \sgpr_{3} \ \sgpr_{2} \ \sgpr_{5}
\ \gnr_{4} \ \sgpr_{6}$ is a successful reduction of $\eoverlap_u$.
\end{Example}

We now show that these two types of rules faithfully simulate the
string negative rule and the simple string positive rule.

\begin{Lemma} \label{snr_simulate}
Let $u$ be a legal string and let $p \in \Pi$. Then $\snr_p$ is
applicable to $u$ iff $\gnr_{\pset{p}}$ is applicable to
$\eoverlap_u$. In this case, $\eoverlap_{\snr_p(u)} =
\gnr_{\pset{p}}(\eoverlap_{u})$.
\end{Lemma}
\begin{Proof}
We have $\snr_p$ is applicable to $u$ iff $u = u_1 pp u_2$ for some
strings $u_1$ and $u_2$ iff $\pset{p}$ is negative in $u$ and the
$\pset{p}$-interval is empty iff $\pset{p}$ is negative in
$\eoverlap_u$ and there is no undirected edge with $\pset{p}$ as
endpoint and there is no directed edge to $\pset{p}$ iff
$\gnr_{\pset{p}}$ is applicable to $\eoverlap_u$.

In this case, $\eoverlap_{\snr_p(u)}$ is obtained from
$\eoverlap_{u}$ by removing vertex $\pset{p}$ and the edges
connected to $\pset{p}$, hence $\eoverlap_{\snr_p(u)}$ is equal to
$\gnr_{\pset{p}}(\eoverlap_{u})$.
\end{Proof}

\begin{figure}
$$
\xymatrix{
u \ar[rr]^{\snr_p} \ar[d]_{\eoverlap} & & \snr_p(u) \ar[d]_{\eoverlap}
\\
\eoverlap_u \ar[rr]^{\gnr_{\pset{p}}} & & \eoverlap_{\snr_p(u)}
}
$$
\caption{A commutative diagram illustrating
Lemma~\ref{snr_simulate}.} \label{fig_comm_snr_simulate}
\end{figure}
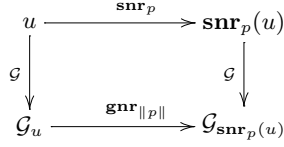

The previous lemma is illustrated as a commutative diagram in
Figure~\ref{fig_comm_snr_simulate}. The next lemma shows that a
similar diagram can be made for the simple string positive rule.

\begin{Lemma} \label{sspr_simulate}
Let $u$ be a legal string and let $p \in \Pi$. Then $\sspr_p$ is
applicable to $u$ iff $\sgpr_{\pset{p}}$ is applicable to
$\eoverlap_u$. In this case, $\eoverlap_{\sspr_p(u)} =
\sgpr_{\pset{p}}(\eoverlap_{u})$.
\end{Lemma}
\begin{Proof}
We have $\sspr_p$ is applicable to $u$ iff $u = u_1 p u_2 \bar p
u_3$ for some strings $u_1$, $u_2$, and $u_3$ with $|u_2| = 1$ iff
$\pset{p}$ is positive in $u$ (or equivalently in $\eoverlap_u$) and
there is exactly one undirected edge $e$ with $\pset{p}$ as endpoint
and there is no directed edge with $\pset{p}$ as endpoint iff
$\sgpr_{\pset{p}}$ is applicable to $\eoverlap_u$.

In this case, $\eoverlap_{\sspr_p(u)}$ is obtained from
$\eoverlap_{u}$ by removing vertex $\pset{p}$, removing all edges
connected to $\pset{p}$, and flipping the sign of the other vertex
of $e$. Hence $\eoverlap_{\sspr_p(u)}$ is equal to
$\gnr_{\pset{p}}(\eoverlap_{u})$.
\end{Proof}

\begin{Example} \label{ex_graph_rules_sim}
In our running example, one can easily verify that the extended
overlap graph of $\sspr_{\bar 6}(u) = 5 \bar 2 4 4 \bar 5 3 \bar 2 b
3 \bar e$ is equal to graph $\sgpr_6(\eoverlap_{u})$ given in
Figure~\ref{fig_gnr_extended_overlapgr}.
\end{Example}

\begin{figure}
\vspace{-0.6cm}
$$
\xymatrix@=10pt{
 & 3^- \ar[d] \\
 & m^- \\
2^- \ar[ur] \ar@{-}[ruu] &  & 4^- \ar@{-}[ll] \ar[lu] \ar@{-}[luu]
}
$$
\caption{The extended overlap graph of $w = b 2 3 4 2 3 4 e$.}
\label{fig_no_sgdr}
\end{figure}
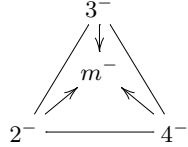

One may be wondering at this point why we have not defined the
simple graph double rule. To this aim, consider extended legal
string $w = b 2 3 4 2 3 4 e$. Note that $\ssdr_{2,3}$ and
$\ssdr_{3,4}$ are applicable to $w$, but $\ssdr_{2,4}$ is not
applicable to $w$. However, this information is lost in
$\eoverlap_{w}$ -- applying the isomorphism that interchanges
vertices $2$ and $3$ in $\eoverlap_{w}$ obtains us $\eoverlap_{w}$
again, see Figure~\ref{fig_no_sgdr}. Thus, given only
$\eoverlap_{w}$ it is impossible to deduce applicability of the
simple graph double rule.

To successfully define a simple graph double rule, one needs to
retain information on which pointers are next to each other, and
therefore different concepts are required. However, this concept
would require that the linear representation of the pointers in an
extended legal string is retained. Hence, string representations are
more natural compared to graph representations.

The next lemma shows that simple marked graphs that are extended
overlap graphs are quite restricted in form. We will use this
restriction in the next section.
\begin{Lemma} \label{lem_dir_acyclic}
Let $u$ be a legal string. Then $\rmundir{\eoverlap_u}$ is acyclic
and transitively closed.
\end{Lemma}
\begin{Proof}
There is a (directed) edge from $p$ to $q$ in
$\rmundir{\eoverlap_u}$ iff the $p$-interval is completely contained
in the $q$-interval of $u$. A nesting relation of intervals is
acyclic and transitive.
\end{Proof}

%$u = u_1 q u_2 p u_3 p u_4 q u_5$ for some strings $u_i$, where each
%occurrence of $p$ and $q$ can be inverted. Consequently, an edge
%from $p$ to $q$ indicates that $p$
%Then $u = u_1 r u_2 q u_3 p u_4 p u_5 q u_6 r u_7$ for some strings
%$u_i$.

\begin{Remark}
We have seen that $\rmpar{\eoverlap_u}$ is the overlap graph of $u$.
Not every graph is an overlap graph -- a characterization of which
graphs are overlap graphs are shown
in~\cite{DBLP:journals/jct/Bouchet94}. Hence, both
$\rmundir{\eoverlap_u}$ and $\rmpar{\eoverlap_u}$ are restricted in
form compared to graphs in general.
\end{Remark}

\section{Characterizing Successfulness}
In this section we characterize successfulness of simple marked
graphs in $S \subseteq \{\gnrset,\sgprset\}$. First we consider the
case $S = \{\gnrset\}$.

\begin{Remark}
In the general (not simple) model, which has different graph pointer
rules and is based on overlap graphs, successfulness in $S$ has been
characterized for those $S$ which includes the graph negative rules
(note that these rules are different from the graph negative rules
defined here) -- the cases where $S$ does not contain the graph
negative rules remain open. \end{Remark}

\begin{Theorem}
Let $\gamma$ be a simple marked graph. Then $\gamma$ is successful
in $\{\gnrset\}$ iff each vertex of $\gamma$ is negative, $\gamma$
has no undirected edges, and $\gamma$ is acyclic.
\end{Theorem}
\begin{Proof}
Since $\rmundir{\gamma} = \gamma$ is acyclic, there is a linear
ordering $(p_1,p_2,\ldots,p_n)$ of the vertices of $\gamma$ such
that if there is an edge from $p_i$ to $p_j$, then $i < j$. The
result now follows by the definition of $\gnr$. In this case, linear
ordering $(p_1,p_2,\ldots,p_n)$ corresponds to a successful
reduction $\varphi = \gnr_{p_{n-1}} \ \cdots \ \gnr_{p_2} \
\gnr_{p_1}$ of $\gamma$.
\end{Proof}

Using Lemma~\ref{lem_dir_acyclic}, more can be said if $\gamma =
\eoverlap_u$ for some legal string $u$.

\begin{Corollary}
Let $\gamma = \eoverlap_u$ for some legal string $u$. Then $\gamma$
is successful in $\{\gnrset\}$ iff each vertex of $\gamma$ is
negative and $\gamma$ has no undirected edges. In this case,
$\gamma$ is the transitive closure of a forest, where edges in the
forest are directed from children to their parents.
\end{Corollary}

Next we turn to the case $S = \{\sgprset\}$.

\begin{Theorem} \label{th_sgpr_char}
Let $\gamma$ be a simple marked graph. Then $\gamma$ is successful
in $\{\sgprset\}$ iff the following conditions hold:
\begin{enumerate}
\item
$\rmpar{\gamma}$ is a (undirected) tree, \\
\item
for each vertex $v$ of $\rmpar{\gamma}$, the degree of $v$ is even
iff
$v$ is negative in $\gamma$, and \\
\item
the graph obtained by replacing each undirected edge in $\gamma$ by
a directed edge from the child to the parent in tree
$\rmpar{\gamma}$ with root $m$ is acyclic.
%
%there is no path of directed edges in $\gamma$ from a vertex $v$ to
%a descendant of $v$ (or $v$ itself) in tree $\rmpar{\gamma}$ with
%root $m$.
\end{enumerate}
\end{Theorem}
\begin{Proof}
Proof sketch. It can be verified that each of both statements hold
iff there is an linear ordering $(p_1,p_2,\ldots,p_n)$ of the
vertices of $\gamma$ such that $p_n = m$, and for each $p_i$ with $i
\in \{1,\ldots,n\}$ the following holds:
\begin{enumerate}
\item
the number of undirected edges from vertices $p_j$ with $j < i$ to
$p_i$ is even iff $p_i$ is positive in $\gamma$,
\\
\item
if $i<n$, then there is exactly one undirected edge between $p_i$
and another vertex $p_j$ with $j > i$, and
\\
\item
there is no directed edge from a vertex $p_j$ to $p_i$ with $j > i$.
\end{enumerate}
In this case, linear ordering $(p_1,p_2,\ldots,p_n)$ corresponds to
a successful reduction $\varphi = \sgpr_{p_{n-1}} \ \cdots \
\sgpr_{p_2} \ \sgpr_{p_1}$ of $\gamma$.
\end{Proof}
%
%Clearly, due to Lemma~\ref{lem_dir_acyclic}, the last requirement is
%redundant in case $\gamma = \eoverlap_u$ for some legal string $u$.

\begin{Example} \label{ex_succ_sgpr}
Consider again extended legal string $u$ of
Example~\ref{ex1_extended_overlapgr} with its extended overlap graph
$\eoverlap_u$ given in Figure~\ref{fig2_extended_overlapgr}. Then by
Theorem~\ref{th_sgpr_char}, $\eoverlap_u$ is \emph{not} successful
in $\{\sgprset\}$, since condition~1 is violated -- $\rmpar{\gamma}$
is not a tree as it has two connected components.

Reconsider now extended legal string $v$ of
Example~\ref{ex1_extended_overlapgr} with its extended overlap graph
$\eoverlap_v$ given in Figure~\ref{fig3_extended_overlapgr}. By
Theorem~\ref{th_sgpr_char}, $\eoverlap_v$ is successful in
$\{\sgprset\}$. According to the proof of
Theorem~\ref{th_sgpr_char}, $(2,4,3,m)$ is a linear ordering of the
vertices corresponding to a successful (graph) reduction $\varphi =
\sgpr_{3} \ \sgpr_{4} \ \sgpr_{2}$ of $\eoverlap_v$. By
Lemma~\ref{sspr_simulate}, this in turn corresponds to a successful
(string) reduction $\varphi'$ of $v$ -- one can verify that we can
take $\varphi' = \sspr_{\bar 3} \ \sspr_{\bar 4} \ \sspr_{2}$.
Moreover, by the proof of Theorem~\ref{th_sgpr_char}, linear
ordering $(4,2,3,m)$ does \emph{not} correspond to a successful
reduction of $\eoverlap_v$ (or of $v$).
\end{Example}

Finally, we consider the case $S = \{\gnrset,\sgprset\}$.

\begin{Theorem} \label{th_gnr_sgpr_char}
Let $\gamma$ be a simple marked graph. Then $\gamma$ is successful
in $\{\gnrset,\sgprset\}$ iff the all of the conditions of
Theorem~\ref{th_sgpr_char} hold, except that in condition 1)
$\rmpar{\gamma}$ is a forest instead of a tree, and in condition 3)
for each tree in the forest we can identify a root, where $m$ is one
such root, such that the graph obtained by replacing each undirected
edge $e$ in $\gamma$ by a directed edge from the child to the parent
in the tree to which $e$ belongs, is acyclic.
\end{Theorem}
\begin{Proof}
Proof sketch. It can be verified that each of both statements hold
iff there is an ordering $(p_1,p_2,\ldots,p_n)$ of the vertices of
$\gamma$ such that for each $p_i$ with $i \in \{1,\ldots,n\}$,
condition 1) holds and \emph{either} conditions 2) and 3) hold in
the proof of Theorem~\ref{th_sgpr_char} \emph{or} there is no edge
(directed or not) between a vertex $p_j$ to $p_i$ with $j > i$.

Again, in this case, linear ordering $(p_1,p_2,\ldots,p_n)$
corresponds to a successful reduction $\varphi$ of $\gamma$ where
the vertices corresponding to roots in forest $\rmpar{\gamma}$
(except $m$) are used in $\gnr$ rules, while the other vertices are
used in $\sgpr$ rules.
\end{Proof}

\begin{Example} \label{ex_succ_gnr_sgpr}
Consider again extended legal string $u$ of
Example~\ref{ex1_extended_overlapgr} with its extended overlap graph
$\eoverlap_u$ given in Figure~\ref{fig2_extended_overlapgr}. By
Theorem~\ref{th_gnr_sgpr_char}, $\eoverlap_u$ is successful in
$\{\gnrset,\sgprset\}$. By the proof of
Theorem~\ref{th_gnr_sgpr_char}, $(6,4,5,2,3,m)$, $(4,6,5,2,3,m)$,
and $(4,5,6,2,3,m)$ are the linear orderings of the vertices that
correspond to successful reductions of $\eoverlap_u$ in
$\{\gnrset,\sgprset\}$. Moreover, in each case vertex $4$
corresponds to the $\gnr_{4}$ rule while the other pointers
correspond to $\sgpr$ rules.
\end{Example}

\section{Discussion}
We have shown that we can partially model simple gene assembly based
on a natural extension of the well-known concept of overlap graph.
The model is partial in the sense that the simple double string rule
does not have graph rule counterpart. Within this partial model we
characterize which micronuclear genes can be successfully assembled
using 1) only graph negative rules, 2) only simple graph positive
rules, and 3) both of these types of rules. These results carry over
to the corresponding simple string pointer rules.

What remains is a graph rule counterpart of the simple double string
rule. However such a counterpart would require different concepts
since the overlap graph or any natural extension does not capture
the requirement that pointers $p$ and $q$ (in the rule) are next to
each other in the string.

\bibliography{../geneassembly}

\end{document}